\documentclass[12pt]{article}
\newdimen\tableauside\tableauside=1.0ex
\newdimen\tableaurule\tableaurule=0.4pt
\newdimen\tableaustep
\def\phantomhrule#1{\hbox{\vbox to0pt{\hrule height\tableaurule width#1\vss}}}
\def\phantomvrule#1{\vbox{\hbox to0pt{\vrule width\tableaurule height#1\hss}}}
\def\sqr{\vbox{%
  \phantomhrule\tableaustep
  \hbox{\phantomvrule\tableaustep\kern\tableaustep\phantomvrule\tableaustep}%
  \hbox{\vbox{\phantomhrule\tableauside}\kern-\tableaurule}}}
\def\squares#1{\hbox{\count0=#1\noindent\loop\sqr
  \advance\count0 by-1 \ifnum\count0>0\repeat}}
\def\tableau#1{\vcenter{\offinterlineskip
  \tableaustep=\tableauside\advance\tableaustep by-\tableaurule
  \kern\normallineskip\hbox
    {\kern\normallineskip\vbox
      {\gettableau#1 0 }%
     \kern\normallineskip\kern\tableaurule}%
  \kern\normallineskip\kern\tableaurule}}
\def\gettableau#1 {\ifnum#1=0\let\next=\null\else
  \squares{#1}\let\next=\gettableau\fi\next}

\begin{document}
NYU-TH-00/09/07
\vspace{20mm}

\begin{center}
{\bf \large Masses of Flavor Singlet Hybrid Baryons}\\
\vspace{20mm}

Olaf ~Kittel\\

{\it Institut f\"ur Theoretische Physik, Universit\"at W\"urzburg, D-97074
W\"urzburg, Germany \\
\vspace{4mm}
and\\}
\vspace{4mm}
Glennys ~R. ~Farrar

{\it Department of Physics, New York University, NY, NY 10003, USA}

\end{center}

\vspace{25mm}

\begin{abstract}

We study the possibility that four iso-singlet baryons
$\Lambda_s(1405)$ $J^P=\frac{1}{2}^{(-)}$, $\Lambda_s(1520)$
$J^P=\frac{3}{2}^{(-)}$, $\Lambda_c(2593)$ $J^P=\frac{1}{2}^{(-)}$ and
$\Lambda_c(2625)$ $J^P=\frac{3}{2}^{(-)}$ are hybrids: three quark one
gluon states $(udsg)$.  We calculate the mass separations of the
candidates, using a degeneracy-lifting hyperfine interaction from an
effective single colored gluon exchange between the constituents.  The
correct ordering of masses is obtained (contrary to the case for the
conventional interpretation as 3 quarks with $L=1$) and the splittings
are plausible.  The parity of these states is not measured, only
assumed to be negative.  In the hybrid picture, the lightest states
are parity even and the parity odd counterparts lie about 300 MeV
higher.  Thus the hybrid ansatz predicts that either the parity of the
$\Lambda(1405)$ etc is positive, or that there are undiscovered
positive parity states about 300 MeV lower.  We also remark that in
this picture, the $H$-dibaryon mass may be around 1.5 GeV.

\end{abstract}
\newpage
\section{Introduction}

The spectrum of hadrons between 1 GeV and 2 GeV has been accurately measured
experimentally \cite{1.0}. In the '60s, Gell-Mann and Zweig \cite{GeNe} 
proposed a systematic classification of the hadrons
into multiplets based on quarks, the fundamental building blocks of matter.
The fact that free quarks had never been observed led Gell-Mann at that time to 
argue that quarks are just a mnemonic device to explain the observed
spectra. The understanding of non-observation of quarks was interpreted by 
\emph{Quantum Chromodynamics} (QCD) in the '70s. 
It is now generally assumed that QCD is the fundamental theory of the strong 
interactions. The quarks that build up the hadrons
are bound together by virtual gluons, the mediators of
the strong force. QCD is an asymptotically free theory; at short
distances (i.e. large momentum transfer) the quark-gluon coupling 
goes to zero and conversely, when quarks are separated from each other by 
large distances, the coupling gets stronger. This effect, known as
``infrared slavery'', leads to quark confinement at low energies. Matter 
consists of color singlets, which are objects with zero total color charge
such as mesons ($q \bar{q}$) or baryons ($qqq$).

QCD predicts other composite particles, color singlet states not only
involving quarks but also ``constituent'' gluons \cite{BaClVi}. The
constituent gluons should not be confused with the virtual gluons.
The constituent gluons contribute with their spin and parity to the
quantum numbers of the hadron.  The phenomenology of these baryons,
called hybrids, was developed by Barnes and Close \cite{BaClo,
BarClos}, Golowich et al. \cite{GoHa} and others. These works use the
bag model and the potential model to suggest that the lightest hybrid
masses are below 2 GeV.  Although there is evidence in favor of some
candidates \cite{Mey}, the hybrids have not yet been found.  One
problem is that the mixing of hybrids with conventional states makes
them hard to detect.  Since the quantum numbers of hybrid states
sometimes coincide with those of known particles, it is also possible
that some particles identified as conventional baryon states could be
hybrids \cite{Karl, FEClose}.

It would be interesting if the spin $\frac{1}{2}$ and $\frac{3}{2}$
isosinglet baryons conventionally identified as an $L=1$ state in the
quark model turned out to be a hybrid baryon state.  A constituent
gluon in a negative parity mode combining with quarks in an $L=0$
orbital ground state can give $J^P=\frac{1}{2}^{(-)}$ and $\frac{3}{2}^{(-)}$
states. The $\Lambda(1405)$ $J^{P}=1/2^{(-)}$, which is still a
mystery in the quark model, could be one such candidate
\cite{Far}. Composed of an up, a down and a strange quark, the
$\Lambda(1405)$ is usually assumed to be a state with orbital
excitation \cite{IsKa}. However there is a major problem in this quark
model interpretation, because it implies the
$\Lambda(1405)$ and its partner, the $\Lambda(1520)$ with
$J^{P}=3/2^{(-)}$, are predicted to be nearly degenerate in mass
\cite{CapIsg}. Even worse, calculations with spin orbit interactions
\cite{CloDal} predict an inversion of the masses. As a second
interpretation, some suggest that the $\Lambda(1405)$ is a bound
state of $\bar{K}$ and $N$ (or a resonance of a pion and a $\Sigma$
\cite{ArMa}). The problem with this interpretation is that 
it only provides a $J^{P}=1/2^{(-)}$ state and not a $J^{P}=3/2^{(-)}$
state. If the $\Lambda(1520)$ is given the conventional interpretation
as an orbital excitation of 3 quarks, another particle with
$J^{P}=1/2^{(-)}$ \cite{Dal} whose mass is close to the $\Lambda(1520)$ is
required. This is ruled out experimentally since this region has
already been explored thoroughly.

In this paper we investigate the possibility that the $\Lambda(1405)$
and its partner $\Lambda(1520)$ are hybrids.  To test this hypothesis,
we calculate the mass splittings of the flavor singlet, octet and
decuplet hybrid baryons.  In section \ref{Wave functions of Hybrids}
we derive the general structure of the hybrid wave function.  In
section \ref{The Interaction Hamiltonian} we introduce the degeneracy
lifting hyperfine interaction. The interaction strength is determined
by an effective 1 gluon exchange between quarks and between a quark
and a gluon.  In section \ref{The hyperfine coupling constant kappa},
the strength of the quark-quark effective hyperfine coupling is
determined from ordinary baryon mass splittings.  In section \ref{Mass
splittings of the hybrid baryons} we calculate the flavor singlet
hybrid mass splittings and discover they have the correct ordering
$m_{3/2}>m_{1/2}$, unlike in the conventional orbital excitation
picture.  We then use the observed flavor singlet splittings to fix
the quark-gluon effective hyperfine coupling.  This allows predictions
to be made for octet and decuplet hybrid baryon mass splittings.  In
sections \ref{The parity of Lambda(1405)} and \ref{A low lying dihyperon ?} two
immplications of this model are discussed -- parity doubling and a
light $H$-dibaryon.  Section \ref{Summary and Conclusion} gives a
summary of our results and conclusions.

\section{\label{Wave functions of Hybrids}
                Wave functions of hybrids}

In this section we study the structure of the hybrid wave 
function. It is determined by the three quark and the 
constituent gluon wave functions. We discuss wave functions 
for quarks in orbital ground states. Because the quarks have to 
obey the Pauli principle, we can show that a systematic
group theoretical classification of all hybrid wave functions is 
possible.

\subsection{\label{Construction of the quark wave functions}
                   Construction of the quark wave functions}

The $SU(N)$ group for color is $SU(3)_C$. A quark transforms like a 
triplet under $SU(3)_C$ because it comes in three colors.
The $SU(N)$ group for spin is $SU(2)_S$. The spin of the quark is 1/2, 
so it is a doublet representation of $SU(2)_S$.
In this paper we consider systems of three quarks having up to three
different quark flavors. The $SU(N)$ group for flavor 
is then $SU(3)_F$, if the mass differences between quarks of different 
flavors are neglected. A quark transforms as a triplet under
$SU(3)_F$, if flavor symmetry is assumed. A single quark is then 
a 18-dimensional representation of the direct product 
group $SU(3)_C \times SU(3)_F \times SU(2)_S$. 
We reduce the direct product of the three quarks in
irreducible representations of $SU(18)$; with the help of Young
tableaux \cite{Sche} we find: 
\begin{eqnarray} \label{SU(18) decomposition}
  {\tableau{1} \atop{\bf 18}} 
  {\times  \atop }
  {\tableau{1} \atop{\bf18}}
  {\times \atop }
  {\tableau{1} \atop{\bf18}} {= \atop }
    {\tableau{3} \atop{\bf 1140}} {+ \atop } 
    {\tableau{2 1} \atop{\bf 1938}} {+ \atop }
    {\tableau{2 1} \atop{\bf 1938}} {+ \atop }
    {\tableau{1 1 1} \atop{\bf 816}} \, .
\end{eqnarray}
We label the resulting irreducible representations by their
dimensions. The completely antisymmetric representation is the {\bf 816}. 
All $qqq$-states that obey the Pauli principle are members of this
multiplet. We begin to decompose the multiplet into representations 
of $SU(3)_C$, $SU(3)_F$ and $SU(2)_S$. We find eight different 
representations of flavor, color and spin (see table \ref{Decomposition}).
\begin{table}
 \[ 
  \begin{tabular}{|l||c|c|c||c|}  \hline
$SU(18)$&$SU(3)_C$ & $SU(3)_F$ & $SU(2)_S$ & \# of these \\ \hline
                                                            \hline 
        & $  { \tableau{1 1 1} \atop\textstyle{\bf 1}}$
        & $  { \tableau{2 1}   \atop\textstyle{\bf  8}}$
        & $  { \tableau{2 1}   \atop\textstyle{\bf 2}}$  &16 \\ \cline{2-5}
        & $  { \tableau{1 1 1} \atop\textstyle{\bf 1}}$
        & $  { \tableau{3}     \atop\textstyle{\bf 10}}$ 
        & $  { \tableau{3}     \atop\textstyle{\bf 4}}$  &40 \\ \cline{2-5} \cline{2-5}
        & $  { \tableau{3}     \atop\textstyle{\bf 10}}$
        & $  { \tableau{2 1}   \atop\textstyle{\bf  8}}$   
        & $  { \tableau{2 1}   \atop\textstyle{\bf  2}}$  &160\\ \cline{2-5}
 $ {\tableau{1 1 1} \atop\textstyle{\bf 816}}$ 
        & $  { \tableau{3}     \atop\textstyle{\bf 10}}$
        & $  { \tableau{1 1 1} \atop\textstyle{\bf  1}}$ 
        & $  { \tableau{3}     \atop\textstyle{\bf  4}}$  & 40\\ \cline{2-5} \cline{2-5}
        & $  { \tableau{2 1}   \atop\textstyle{\bf  8}}$
        & $  { \tableau{2 1}   \atop\textstyle{\bf  8}}$  
        & $  { \tableau{2 1}   \atop\textstyle{\bf  2}}$  &128\\ \cline{2-5}
        & $  { \tableau{2 1}   \atop\textstyle{\bf 8}}$
        & $  { \tableau{2 1}   \atop\textstyle{\bf 8}}$  
        & $  { \tableau{3}     \atop\textstyle{\bf 4}}$  &256\\ \cline{2-5}
        & $  { \tableau{2 1}   \atop\textstyle{\bf 8}}$
        & $  { \tableau{3}     \atop\textstyle{\bf 10}}$   
        & $  { \tableau{2 1}   \atop\textstyle{\bf  2}}$  &160\\ \cline{2-5}
        & $  { \tableau{2 1}   \atop\textstyle{\bf  8}}$
        & $  { \tableau{1 1 1} \atop\textstyle{\bf  1}}$
        & $  { \tableau{2 1}   \atop\textstyle{\bf  2}}$  & 16\\ \hline
   \end{tabular}
  \]
 \caption{\label{Decomposition}Decomposition of $SU(18)$ in
                  representations of color, flavor and spin.}
\end{table}
We can check if we have listed all possible decompositions. By
counting the numbers of the states in table \ref{Decomposition}
we find 816, so we are consistent.

The gluon is in the color octet representation of $SU(3)_C$. The hybrid
has to be a color singlet. Therefore, the quarks have to be in the 
complex conjugate representation. This is again an octet, so only color octet
$qqq$-states are of interest here. We can discard all other states.
We will distinguish the color octets in the lower half of 
table \ref{Decomposition} by the short hand
$^{spin}flavor$: $ \bf{ ^2 8}, \:^4 8,\: ^2 10,\: ^2 1$. 
There are altogether 70 color octet states forming a $SU(6)$ 
representation of $SU(3)_F \times SU(2)_S$ with mixed symmetry:
\begin{eqnarray}  \label{SU(6) reduction}
  {\tableau{1} \atop {\bf 6}} 
  {\times  \atop }
  {\tableau{1} \atop{\bf6}}
  {\times \atop }
  {\tableau{1} \atop{\bf6}} {= \atop }
    {\tableau{3} \atop{\bf 56}} {+ \atop } 
    {\tableau{2 1} \atop{\bf 70}} {+ \atop }
    {\tableau{2 1} \atop{\bf 70}} {+ \atop }
    {\tableau{1 1 1} \atop{\bf 20}} \, .
\end{eqnarray}

Young tableaux of the form 
$\tableau{2 1}$ are called mixed symmetric because
the wave function which is assigned to the  tableaux $\tableau{2 1}$
is neither completely symmetric nor antisymmetric. But the wave function
still has defined symmetry properties under the exchange of two quarks.
We choose them to be the first and second quark and will use this 
convention throughout. In the wave function the first and second quark 
can be either coupled in a symmetric or antisymmetric way, shown below.
 \begin{eqnarray}
  \begin{array}{lccr}
  &\tableau{1} \times \tableau{1}&=&\tableau{2}+\tableau{1 1}\\
   SU(3): & {\bf3} \times {\bf3} &=& {\bf6 }+ {\bf\bar{3}} \\
   SU(2): & {\bf2} \times {\bf2} &=& {\bf3} + {\bf1}
  \end{array}
 \end{eqnarray}
We label the wave function $\varphi$ by a subscript $MS$ 
for mixed symmetric states ($\varphi_{MS}$) and $MA$ for 
mixed antisymmetric states ($\varphi_{MA}$) to indicate these 
important symmetry properties of the first and second 
quark. Because the choice of this basis is arbitrary, we give the
resulting form of the wave function explicitly in appendix
\ref{Definitions of the mixed symmetry functions}. 
We can build a totally symmetric (totally antisymmetric) 
function $\phi_{S(A)}$ out of two mixed symmetric functions 
$\varphi_{MS(A)}$ and $\varphi_{MS(A)}'$ in the
following way \cite{Closebook}: 
 \begin{eqnarray}
      \phi_S &=& \displaystyle
                \frac{1}{\sqrt{2}}
      (\varphi_{MS}\varphi_{MS}' + \varphi_{MA}\varphi_{MA}') \\
      \phi_A &=& \displaystyle
                \frac{1}{\sqrt{2}}
      (\varphi_{MS}\varphi_{MA}' - \varphi_{MA} \varphi_{MS}' ).
  \end{eqnarray}
The factor $1/\sqrt{2}$ is normalization.
We can also form mixed symmetric functions $\phi_{MS}$ 
and $\phi_{MA}$:
  \begin{eqnarray}
      \phi_{MS} &=& \frac{1}{\sqrt{2}}
      (-\varphi_{MS}\varphi_{MS}' + \varphi_{MA}\varphi_{MA}') \\
      \phi_{MA} &=& \frac{1}{\sqrt{2}}
      (\varphi_{MS}\varphi_{MA}' + \varphi_{MA} \varphi_{MS}' ).
 \end{eqnarray}
We are now able to write down the structure of the four color octet wave
functions, that lie in the 816 dimensional representation 
of $SU(18)$ \cite{Closebook}. 
$f,\: s,\: c$ denote
the flavor, spin and the color three quark wave functions, respectively. 
\begin{eqnarray} \label{48 wave function}
 {\bf^4 8}&:& \displaystyle
        \frac{1}{\sqrt{2}}
                    \big(
                        c_{MS}f_{MA}- c_{MA}f_{MS}
                    \big)  s_{S} \\ \label{210 wave function}
 {\bf^2 10}&:&   \displaystyle
         \frac{1}{\sqrt{2}}(c_{MS}s_{MA}-c_{MA}s_{MS})f_{S}\\ 
 {\bf^2 8}&:& \displaystyle
         \frac{1}{2}
                \Big[ 
                       c_{MS}(f_{MA}s_{MS}+f_{MS}s_{MA})
                      -c_{MA} (f_{MA}s_{MA}-f_{MS}s_{MS} )
                \Big]\\ \label{The singlet wave function}
 {\bf^2 1}&:& \displaystyle 
        \frac{1}{\sqrt{2}}(c_{MA}s_{MA} + c_{MS}s_{MS})f_{A}.
  \end{eqnarray}

\subsection{\label{Construction of hybrid wave functions}
                   Construction of hybrid wave functions}

We are interested in the flavor
singlet sector of the $qqq$ color octet. This is
the wave function from eqn. \ref{The singlet wave function}:
  \begin{eqnarray}
  {\bf^2 1}&:& \displaystyle 
        \frac{1}{\sqrt{2}}(c_{MA}s_{MA} + c_{MS}s_{MS})f_{A}.
        \end{eqnarray}
This wave function carries a suppressed color index that gets
contracted with the color index of the gluon when forming the
$qqqg$-state. Only the spin combination is left, as the gluon is
flavorless.

The free gluon is a massless particle with two helicity states.
In confining the gluon inside the hybrid, it is assumed 
\cite{BarClos, GoHa} that the gluon 
gains mass and with that a third degree of freedom, the spin 0 state.
In the limit of zero coupling, the problem can be modeled by the
familiar problem
in classical electrodynamics to find the cavity normal modes for a massless
photon field in a rigid spherical cavity \cite{TaDo}. The confined field is
classified by TM or TE modes, with defined total angular momentum $J$ and 
parity $P$:
 \begin{eqnarray}
   \begin{array}{ccl}
      TE &:& J^P=1^{+}, 2^{-}, \dots \\
      TM &:& J^P=1^{-}, 2^{+}, \dots \; .
   \end{array}
   \end{eqnarray}
(With the assumed boundary conditions, and standard bag model
parameters, the TM$(1^{-})$ mode is  more energetic than the lowest
energy eigenmode 
TE$(1^{+})$.) We can construct the quantum numbers of the hybrid state with 
these $J^P$ values.
With the $J=1$ gluon mode and the flavor singlet three quark state $J=1/2$,
we can form two $qqqg$-states, one with spin 1/2 and one with spin 3/2.
In the latter case, gluon spin and $qqq$ spin are aligned, in the former,
they are opposite. This will give the main contribution to the mass
splittings between these two states.

\section{\label{The Interaction Hamiltonian}
                The interaction Hamiltonian}

In this section we discuss the Hamiltonian that we
use to determine mass splittings of hybrid baryons. 
The hyperfine interaction Hamiltonian $V_{hyp}$ can 
be divided into two parts:
\begin{eqnarray} \label{Vhyp}
 V_{hyp} & = & V_{qq} + V_{qg}.
\end{eqnarray}
The first part $ V_{qq}$ includes the interaction only between the quarks.
The second part $ V_{qg}$ contains the interaction between  the quarks
and the  constituent gluon.

The  interaction $ V_{qq}$ between the three quarks is proportional 
to \cite{GoHa}
\begin{eqnarray}
      V_{qq} & \propto &  - \sum_{i<j}S^i\cdot S^jF^i \cdot F^j.
\end{eqnarray}
$S^i$ and $F^i$ are the spin and color matrices for the $i$th 
quark. Quarks are in the fundamental representation of $SU(2)_S$ and 
$SU(3)_C$. So the spin and color matrices for a quark are defined by 
\begin{eqnarray}
          S =  \frac{1}{2}\sigma, \, \, \,
          F =  \frac{1}{2}\lambda. 
\end{eqnarray}
$\sigma$ are the Pauli and $\lambda$ are the Gell-Mann matrices,
given in appendix \ref{The Pauli and the Gell-Mann matrices}.
We define the dot products $S\cdot S$ and $F \cdot F$
in appendix \ref{Casimir operators}. This effective hyperfine 
interaction can be interpreted as the dominant contribution from one-gluon
exchange between quarks inside the baryon with a radius $r \approx 1\, \, fm$. 

The effective one-gluon ansatz leads to the quark-gluon interaction \cite{GoHa}
\begin{eqnarray}
       V_{qg} & \propto & -  \sum_{i}S^i\cdot S^gF^i\cdot F^g,
\end{eqnarray}
where $S^g$ and $F^g$ are the spin and color matrices for the
gluon. 
In ref. \cite{BuFaPu} a similar system is studied: baryons coupled to
a gluino ($\tilde{g}$), the supersymmetric fermionic partner of the
gluon, and many of those techniques can be used here.
We however assume different coupling strengths\footnote{The effective 
coupling strength in \cite{BuFaPu} also depends on the 
radius $r$ of the baryonic state. However,
bag model calculations \cite{thesis} show that the radius changes only of 
the order of a few percent. Because we assume an effective coupling 
with an accuracy about ten percent, this effect is negligible, and we use
radius independent couplings.}
between the light
quarks, a light and a heavy quark, a gluon and a light quark 
and a gluon and a heavy quark which breaks the $SU(3)_F$ symmetry. We
label them $\kappa$, $\kappa_i$, 
$\kappa_g$ and $\kappa_{ig}$ respectively, with $i$ the index for the heavy
quark $(s, c, b)$. $\kappa$ and $\kappa_i$ can be
determined from ordinary baryon mass splittings. This is done in
section \ref{The hyperfine coupling constant kappa}.

Our final interaction Hamiltonian is
\begin{eqnarray} \label{the interaction Hamiltonian}
      V_{hyp} & = & -\sum_{i<j}\kappa_{ij}S^i\cdot S^jF^i\cdot F^j
          -\sum_{i}\kappa_{ig} S^i\cdot S^gF^i\cdot F^g.
\end{eqnarray}

The hierarchy of the $\kappa$'s can be guessed by the form
of the ``Fermi-Breit-interaction'' in QCD for single gluon exchange
given by \cite{FEClose}:
  \begin{eqnarray}
   H_I \propto \displaystyle
            \frac{F^i\cdot F^jS^i\cdot S^j}{m_i \;m_j}.
 \end{eqnarray}
The effective coupling $\kappa$ would therefore be expected to be 
inversely proportional to the product of the masses of the interacting 
particles:
\begin{eqnarray} \label{kappa mass relation}
  \frac{ \kappa_i}{\kappa_j}= \frac{m_j}{m_i}.
\end{eqnarray}
We assume that the hyperfine coefficient is the product of
the color magnetic moments. In going from $\kappa$ to $\kappa_g$
we are replacing a light quark color magnetic moment with a gluon
magnetic moment. Since this is the same replacement independent of quark
flavor, we would expect that
 \begin{eqnarray} \label{kappa relation}
       \frac{\kappa}{\kappa_{g}}= 
    \frac{\kappa_s}{\kappa_{sg}}=  
    \frac{\kappa_c}{\kappa_{cg}}.
 \end{eqnarray}
This reduces the number of parameters in the fit of hybrid
baryons and makes it more predictive.

\subsection{\label{Quark-Quark Interactions}
                Quark-quark interactions}

In this section we calculate the matrix elements of the interaction 
Hamiltonian $V_{qq}$ (see eqn. \ref{the interaction Hamiltonian}).
The quark-quark interaction is given by 
\begin{eqnarray}
      V_{qq} &=& 
                  - \sum_{i<j} \, \kappa_{ij} S^i\cdot S^j
                  F^i\cdot F^j. 
\end{eqnarray}
$ \kappa_{ij}$ is the effective coupling between the constituent 
quarks. The matrix element $E_{hyp}=~<V_{qq}>$ has to be evaluated in 
the quark wave function, which has special symmetry properties.

In section \ref{Construction of the quark wave functions} we discussed
the quark wave function for exact flavor symmetry. We will now
include flavor breaking but will assume that isospin is a good
symmetry. If we consider $SU(3)_F$ breaking, the wave function of a
baryon with two light quarks ($u$ or $d$) has the structure 
$\Psi = \Psi^A +\epsilon \Psi^{12} $. The part $\Psi^A$ is
antisymmetric under interchange of any two quarks. The part $\Psi^{12}$ 
is antisymmetric only under interchange of quark 1 and 2 in the 
basis (see appendix \ref{Definitions of the mixed symmetry
functions}) in which the light quarks are chosen to be 1 and 2. 
When $SU(3)_F$ is unbroken, $\epsilon$ goes to zero. The
measure for flavor breaking is the mass difference between the heavy
and the light quark, so $\epsilon$ is proportional to 
$ m_h - m_l$. To evaluate matrix elements of an operator $O$ 
in  state $\Psi$, we decompose $O$ into a
sum of operators which are either totally antisymmetric  under interchange of
any pair of quarks $(O^A)$ or totally antisymmetric under interchange
of the first and second quark $(O^{12})$: 
\begin{eqnarray}
<O>&=& <O^A>+<O^{12}>. 
\end{eqnarray}
According to these arguments, we decompose the quark-quark interaction 
as follows:
\begin{eqnarray}\label{qq interaction}
      V_{qq} &=& 
                   - \kappa_3 \, \underbrace{\sum_{i<j} S^i\cdot S^j
                                             F^i\cdot F^j
                                            }_{term \, 1}
                   - (\kappa - \kappa_3) \, 
                                 \underbrace{S^1\cdot S^2
                                             F^1\cdot F^2
                                            }_{term \, 2}.
\end{eqnarray}
The first term is completely symmetric. Its  evaluation  has been done 
in \cite{BuFaPu}. The authors found with the help of permutation operators 
for exact flavor symmetry:
 \begin{eqnarray} \label{Expectation value of OA}
  <O^A> &=& < \sum_{i<j} S^i\cdot S^j F^i\cdot F^j>
         =  \frac{21}{16}  - \frac{1}{8}C_C^{qqq}
          - \frac{1}{4}C_F^{qqq} - \frac{1}{12}C_S^{qqq}. 
 \end{eqnarray}
The expectation value is written 
in terms of Casimir operators $C$ (see Appendix \ref{Casimir operators} 
for definition and values in various representations).
There is another nice evaluation of (\ref{Expectation value of OA})
by Jaffe \cite{Jaffe}, involving the $SU(6)$ Casimir operator $C_6$ 
of color and spin:
\begin{eqnarray}
 <O^A> &=& < \sum_{i<j} S^i\cdot S^j F^i\cdot F^j>
         =  \frac{1}{32}C_6^{qqq} - \frac{1}{12}C_S^{qqq} 
         - \frac{1}{8}C_C^{qqq} - \frac{3}{2}.
  \end{eqnarray}    
The definition and values of $C_6$ in various representations may be
found in the appendix of \cite{Jaffe2}.         
We give the values for  $ <\sum_{i<j} S^i\cdot S^j F^i\cdot F^j> $ 
for the color octet three quark states in the first part of table 
\ref{The quark-quark interactions for color octets}.
\begin{table}[h]
 \[ 
  \begin{array}{|c||c|c|c|c|} \hline
SU(3)_C & {\bf8} & {\bf8} & {\bf8} &  {\bf8} \\ \hline
SU(3)_F & {\bf8} & {\bf8} & {\bf10}&  {\bf1} \\ \hline
SU(2)_S & {\bf2} & {\bf4 }& {\bf2} &  {\bf2} \\ \hline \hline
<O^A>   &  1/8   & -1/8   &  -5/8  &     7/8 \\ \hline
  \end{array} \;\;\;\;\;\;\;\;\;
  \begin{array}{|c||c|c|c|c|} \hline
SU(3)_C & {\bf\bar{3}} & {\bf\bar{3}} & {\bf6} & {\bf6} \\ \hline
SU(2)_I &    {\bf1}    &    {\bf3}    & {\bf1} & {\bf3} \\ \hline
SU(2)_S &    {\bf1}    &    {\bf3}    & {\bf3} & {\bf1} \\ \hline \hline
<O^{12}>&   1/2        & -1/6         &   1/12 &  -1/4  \\ \hline
  \end{array}
  \]
  
  \caption{\label{The quark-quark interactions for color octets}
          Expectation values of $ O^A=\sum_{i<j} S^i\cdot S^j F^i\cdot F^j$ 
                 and  $ O^{12}=S^1\cdot S^2 F^1\cdot F^2$ }

\end{table}

The second term of eqn. \ref{qq interaction} is $O^{12}$.
This term is only symmetric under interchange of quark 1 and 2. 
It has to be evaluated in the wave function with broken flavor 
symmetry. Using isospin rather than flavor, we can modify 
eqn. \ref{Expectation value of OA} for this case and find: 
\begin{eqnarray} \label{Expectation value of O12}
 <O^{12}> = <S^1\cdot S^2F^1\cdot F^2> =
     \frac{2}{3}-\frac{1}{8} C_C^{12} -
                 \frac{1}{4} C_I^{12} -
                 \frac{1}{12}C_S^{12}.
\end{eqnarray}
$C_{C,I,S}^{12}$ are the 1-2 diquark Casimir operators for color, isospin
and spin, respectively. We give the values for 
$ <S^1\cdot S^2 F^1\cdot F^2> $ for all possible antisymmetric
representations in flavor, color and isospin of the diquark in the second 
part of table \ref{The quark-quark interactions for color octets}.

\subsection{\label{Quark-gluon interactions}
                   Quark-gluon interactions}

In this section the quark-gluon interaction is evaluated. The 
quark-gluon interaction is given by (\ref{the interaction Hamiltonian})
  \begin{eqnarray}
     V_{qg} &=& 
                - \sum_i \kappa_{ig} S^i\cdot S^gF^i\cdot F^g. 
  \end{eqnarray}
$\kappa_{ig} $ is the effective coupling between the 
constituent gluon and the quarks. According to the arguments
made in section \ref{Quark-Quark Interactions}, the quark-gluon 
interaction has to be decomposed as follows:
 \begin{eqnarray} \label{quark-gluon decomposition}
     V_{qg} &=& 
                - \kappa_{3g}\underbrace{
                                    \sum_i S^i\cdot S^gF^i\cdot F^g
                                        }_{term \, 1} \nonumber \\
            &&  - (\kappa_{g}-\kappa_{3g})\underbrace{
                    (S^1\cdot S^gF^1\cdot F^g+S^2\cdot S^gF^2\cdot F^g)
                                                    }_{term \, 2}. 
  \end{eqnarray}  
The first term is completely symmetric under interchange of any 
pair of quarks. The second term is symmetric only under interchange of 
quark 1 and 2.

The computation of the matrix elements of the first term
has already been done for the light flavor octets and  
decuplets by Barnes and Close \cite{BarClos}. We list our new results for 
the flavor singlet and for completeness, the Barnes and Close results for 
the light octets and decuplets, in table \ref{Values for sumiSiSgFiFg}.
For the benefit of the reader we give a sample computation for the 
flavor singlet in Appendix
\ref{Sample calculation for the flavor singlet}.
\begin{table}[h]
 \[
   \begin{tabular}{|c||c|c|c||c|c||c|c||c|c||} \hline
 $^{spin} flavor$ & \multicolumn{3}{c||}{$\bf^48$} &
                    \multicolumn{2}{c||}{$\bf^210$}&
                    \multicolumn{2}{c||}{$\bf^28$}&
                    \multicolumn{2}{c||}{$\bf^21$} \\ \hline
  total J & 5/2&3/2&1/2&3/2&1/2&3/2&1/2&3/2&1/2  \\ \hline \hline
$<\sum_{i} S^i\cdot S^gF^i\cdot F^g>$&   
    -3/2& 1&5/2&0&0&-1/2&1&-1 &2\\ \hline \hline
        &   \multicolumn{3}{c||}{I=1}&
            \multicolumn{2}{c||}{I=1}&
            \multicolumn{2}{c||}{I=1}&
            \multicolumn{2}{c||}{I=0} \\ \cline{2-10}
$<S^1\cdot S^gF^1\cdot F^g>$
  & -3/8&1/4&5/8&-1/4&1/2&-1/4&1/2&0&0 \\  \cline{2-10}
    &       \multicolumn{3}{c||}{I=0}&
            \multicolumn{2}{c||}{}&
            \multicolumn{2}{c||}{I=0}&
            \multicolumn{2}{c||}{} \\ \cline{2-4}  \cline{7-8}         
    &-5/8& 5/12&25/24&\multicolumn{2}{c||}{} &
                 0&0& \multicolumn{2}{c||}{}\\ \hline
   \end{tabular}
  \]
 \caption{\label{Values for sumiSiSgFiFg}
                 Values for $<\sum_{i} S^i\cdot S^gF^i\cdot F^g>$
                 and $<S^1\cdot S^gF^1\cdot F^g>$}                 
\end{table}

For the light color octets and light decuplets,
term 2 does not contribute at all because $\kappa_{g}=\kappa_{3g}$.
The evaluation of this term is more delicate for hybrids which contain
one heavy quark $i$. 
As we assume that flavor is broken, but that isospin is still a good
symmetry, isospin singlet $(I=0)$ and isospin triplet $(I=1)$ hybrid
baryons arise\footnote{For the quark content $uds$, 
the hypercharge is zero. Thus, the $I_3$ equals the electric charge 
of the hybrids. We could imagine to measure the isospin experimentally
by measuring the charge of the hybrids. For isotriplets we expect three
mass degenerate hybrids with charge -,0,+ and for the isosinglet a single
hybrid with charge 0 but different mass.}.
For the hybrid states we find
\begin{eqnarray}
  {\bf ^21}(I=0), \;\;\;\;\;{\bf ^210}(I=1), \;\;\;\;\;
  {\bf ^48}(I=0,1),\;\;\;\;\; {\bf ^28}(I=0,1).
\end{eqnarray}
For $^{spin}flavor$ fixed, hybrids within isospin multiplets have the same
mass, and hybrids between isospin multiplets have different mass. In order
to compute term 2 we use the symmetry under interchange of quark 1 and 2:
\begin{eqnarray} \label{symmetry 12}
          <S^1\cdot S^gF^1\cdot F^g> &=& <S^2\cdot S^gF^2\cdot F^g>.
\end{eqnarray}
We can write
\begin{eqnarray} 
  \lefteqn{
           <S^1\cdot S^gF^1\cdot F^g>
          }      \nonumber \\
               &=&  \frac{1}{4} 
                       \left[        
                            (S^{1}+S^{g})^2
                           -(S^{1})^2-(S^{g})^2 
                        \right] 
                       \left[        
                            (F^{1}+F^{g})^2
                           -(F^{1})^2-(F^{g})^2 
                        \right]   \nonumber   \\     
               &=&  \frac{1}{4} 
                       \left[       
                              (S^{1}+S^{g})^2 -\frac{11}{4} 
                      \right]    
                       \left[     
                              (F^{1}+F^{g})^2 -\frac{13}{3} 
                       \right].  \label{xy}
\end{eqnarray}
What remains is to know the spin and color representations of the 
gluon--first quark state ($S^1+S^g$ and $F^1+F^g$) of each hybrid baryon. 
In order to expand the color and spin wave function of
each hybrid (with the help of $SU(2)$ \cite{1.0} and $SU(3)$ \cite{Clebsch} 
Clebsch Gordan coefficients) 
into  gluon--first quark and second quark--third quark color and spin
wave functions, we need to know the spin and color representations of 
the diquark. The 1-2 symmetric part of each hybrid wave function has to 
fulfill two constraints. Firstly, the flavor part must be
$f_{MS}$ (or $f_S$) for isotriplets and $f_{MA}$ (or $f_A$) for isosinglets.
Secondly, the wave function must be antisymmetric under interchange of quark
1 and 2. The wave functions for the $\bf ^48$ hybrids result immediately
from (\ref{48 wave function}):
\begin{eqnarray}
  {\bf ^48}(I=1)&:& f_{MS}c_{MA}s_S, \\
  {\bf ^48}(I=0)&:& f_{MA}c_{MS}s_S.
\end{eqnarray}
For the other hybrids (see (\ref{210 wave function}) -
(\ref{The singlet wave function})), the wave function of each hybrid 
may be a linear combination ($lc$) of the following functions:
\begin{eqnarray}
  {\bf ^21}(I=0)&:& lc\;\;of\;\; f_Ac_{MS}s_{MS} \;\;and \;\; f_Ac_{MA}s_{MA}, \\
 {\bf ^210}(I=1)&:& lc\;\;of\;\; f_Sc_{MS}s_{MA} \;\;and \;\; f_Sc_{MA}s_{MS}, \\
  {\bf ^28}(I=1)&:& lc\;\;of\;\; f_{MS}c_{MS}s_{MA}\;\;and \;\; f_{MS}c_{MA}s_{MS}, \\
  {\bf ^28}(I=0)&:& lc\;\;of\;\; f_{MA}c_{MS}s_{MS}\;\;and \;\; f_{MA}c_{MA}s_{MA}. 
\end{eqnarray}
All the above parts of the wave functions are eigenfunctions 
of the operator
\begin{eqnarray}
 A & = & -  (\kappa - \kappa_3) \,S^1 \cdot S^2F^1 \cdot F^2 
         - 2(\kappa_g-\kappa_{3g})\, S^1 \cdot S^gF^1 \cdot F^g,
\end{eqnarray}
which is the complete 1-2 symmetric part of our interaction Hamiltonian 
$V_{hyp}$ (\ref{Vhyp}). We assume that the diquark is in a state
in which the energy is minimal, i.e., in which $A$ is minimized.
The wave functions with minimal energy are:
\begin{eqnarray}
  {\bf ^21}(I=0)&:&  f_Ac_{MA}s_{MA}, \\
 {\bf ^210}(I=1)&:&  f_Sc_{MA}s_{MS}, \\
  {\bf ^28}(I=1)&:&  f_{MS}c_{MA}s_{MS}, \\
  {\bf ^28}(I=0)&:&  f_{MA}c_{MA}s_{MA}. 
\end{eqnarray}
The eigenvalues of $A$ were computed with the estimate values (in MeV)
$\kappa=290$, $\kappa_3 = 180$, $\kappa_g=60$ and $\kappa_{3g}=40$.
We list the resulting values for $<S^1\cdot S^gF^1\cdot F^g>$ 
in table \ref{Values for sumiSiSgFiFg}.
Values for $<S^1\cdot S^2F^1\cdot F^2>$ can be taken from the second 
part of table \ref{The quark-quark interactions for color octets}.

\subsection{\label{The hyperfine coupling constant kappa}
                The hyperfine coupling constant $\kappa$}

We fit the mass splittings of ordinary baryons to find values for
$\kappa$, $\kappa_s$ and $\kappa_c$. 
For the fit we use the masses of the isospin multiplets given in table 
\ref{The particles used to determine kappai}.
We use the average mass of each isospin multiplet.
(The members within isospin multiplets stay degenerate
since we assume exact isospin symmetry and neglect electroweak 
interactions, which are of the order of $0.5\%$).
The listed isospin multiplets $\Sigma_i^*$, $\Sigma_i$ and $\Lambda_i$
are labeled by the flavor index of the heavy quark $i=s,c,b$.
\begin{table}[h]
 \[
   \begin{tabular}{|r|c|c|c|c|c|c|c|c|c|} \hline
  $\kappa_i$&  particle&spin&isospin&flavor&content&
          $\Delta M/MeV$&fit&$\%$&$\kappa_i/MeV$\\ \hline
   $\kappa$ &$\Delta(1232)$& 3/2&3/2& {\bf10}&qqq& $\Delta-N=293$&293&&293   \\
            &   $N(939)$   & 1/2&1/2& {\bf8} &qqq& && &               \\ 
                              \hline
  $\kappa_s$&$\Sigma_s^*(1385)$&3/2&1&{\bf10}&qqs&$\Sigma_s^*-\Sigma_s=192$&182&5&182\\ 
            &$\Sigma_s(1193)$& 1/2&1& {\bf8} &qqs&$\Sigma_s-\Lambda_s=77 $&74&4&\\  
            &$\Lambda_s(1116)$& 1/2&0& {\bf8}&qqs&$\Sigma_s^*-\Lambda_s=269 $&256&5&\\
                              \hline
  $\kappa_c$&$\Sigma_c^*(2520)$&3/2&1&{\bf10}&qqc&$\Sigma_c^*-\Sigma_c=65 $&60&8&60\\ 
            &$\Sigma_c(2455)$& 1/2&1& {\bf8 }&qqc&$\Sigma_c-\Lambda_c=170 $&155&9& \\ 
            &$\Lambda_c(2285)$& 1/2&0& {\bf8}&qqc&$\Sigma_c^*-\Lambda_c=235 $&215&8& \\  
                              \hline      
  $\kappa_b$&$\Sigma_b^*(?)$& 3/2&1& {\bf10}&qqb&$\Sigma_b^*-\Sigma_b=? $&18&20&18\\ 
            &$\Sigma_b(?)$& 1/2&1& {\bf8 } &qqb&$\Sigma_b-\Lambda_b=? $&183&20&\\ 
            &$\Lambda_b(5640)$& 1/2&0& {\bf8}&qqb&$\Sigma_b^*-\Lambda_b=? $&201&20&\\ 
                             \hline
   \end{tabular}
  \]
 \caption{\label{The particles used to determine kappai}
                The particles used to determine $\kappa_i$}
\end{table}
If we assumed exact $SU(3)_F \times SU(2)_S$ symmetry, 
the $\Sigma_i^*$, $\Sigma_i$ and $\Lambda_i$ (for $i$ fixed) would be
members of the totally symmetric 56-dimensional representation.
(compare eqn. (\ref{SU(6) reduction})).
By switching on the mass-difference between the heavy quark $i$ and the 
light quarks $q$ we break $SU(3)_F$. The 
values of $\kappa$, $\kappa_s$ and $\kappa_c$ can thus be determined 
by the experimentally observed mass separations of the isospin multiplets. 
Calculations for mass splittings of this kind are summarized in the book 
by Close \cite{Closebook}, 
p. 387. We obtain the same values of the matrix elements (defined in eqns. 
\ref{Expectation value of OA} and \ref{Expectation value of O12}) 
for the quark-quark interaction
(eqn. \ref{qq interaction})
\begin{eqnarray}
      V_{qq} &=& 
                   - \kappa_3 \, \sum_{i<j} S^i\cdot S^j
                                             F^i\cdot F^j
                   - (\kappa - \kappa_3) \, 
                                 S^1\cdot S^2
                                             F^1\cdot F^2,
\end{eqnarray}
and list the values for $E_{hyp}=<V_{qq}>$ in table \ref{Values for E_hyp}.
\begin{table}[h]
 \[
   \begin{tabular}{|c|c|c|c|c|c|c|} \hline
particle & flavor & \multicolumn{2}{|c|}{diquark} & 
      $<\sum_{i<j} S^i\cdot S^j F^i\cdot F^j>$& 
      $ <S^1\cdot S^2 F^1\cdot F^2>          $ & $E_{hyp}$\\ \hline
    $\Sigma_i^*$ & $ {\tableau{3} \atop\textstyle{\bf 10}}$ 
   &$ C_S^{12}\{ {\bf3}\}$ & $  C_I^{12}\{ {\bf3}\}$ & -1/2& -1/6&  
      $\frac{1}{6} \kappa + \frac{1}{3} \kappa_3 $
        \\  \hline
    $\Sigma_i$   & $ {\tableau{2 1} \atop \textstyle{ \bf8}}$
   &$ C_S^{12}\{ {\bf3}\}$ & $  C_I^{12}\{ {\bf3}\}$ & 1/2& -1/6&    
       $\frac{1}{6} \kappa - \frac{2}{3}\kappa_3$
       \\  \hline
   $\Lambda_i$   & $ {\tableau{2 1} \atop \textstyle{\bf 8}}$
   &$ C_S^{12}\{ {\bf1} \}$ & $  C_I^{12}\{ {\bf1}\}$ & 1/2& 1/2 &
        $-\frac{1}{2} \kappa $  \\ \hline
   \end{tabular}
  \]
 \caption{\label{Values for E_hyp}
                 Values for $E_{hyp}$}                 
\end{table}

With these values at hand, we find for the mass splittings:

\begin{eqnarray}
  E(\Sigma_i^*) - E(\Sigma_i)&=& \kappa_i,   \\ 
 E(\Sigma_i) - E(\Lambda_i) &=& \frac{2}{3}\kappa - \frac{2}{3}\kappa_i, \\ 
  E(\Sigma_i^*) - E(\Lambda_i)&=& \frac{2}{3}\kappa + \frac{1}{3}\kappa_i.
\end{eqnarray}

With increasing mass of the heavy quark $i$, $\kappa_i$ goes 
down (\ref{kappa mass relation}) so that the $\Sigma_i^*-\Sigma_i$
splittings go down, the $\Sigma_i -\Lambda_i$ splittings go
up and the  
$\Sigma_i^*- \Lambda_i$ splittings go down.
This effect can be observed in table 
\ref{The particles used to determine kappai}, where we included 
the results for our $\kappa_i$ fits on the right side.

The masses of the $\Sigma_b^*$ and $\Sigma_b$ have not yet been 
measured. We can predict the order of their mass splittings. We use 
relation (\ref{kappa mass relation})
\begin{eqnarray}
   \kappa_b= \frac{m_c}{m_b} \kappa_c
\end{eqnarray}
to find an estimate value for $\kappa_b$.
With values $m_c=(1.25 \pm 0.15)$ GeV and $m_b=(4.25 \pm 0.15)$
 GeV from  \cite{1.0}, we find e.g. for:
\begin{eqnarray}
 \Sigma_b^*(?) -\Sigma_b(?) &=& (18 \pm 4)   \, MeV = \kappa_b.
\end{eqnarray}
The errors result from the uncertainty of the quark masses.
Errors arising due to  assumption 
(\ref{kappa mass relation}) are not included.

\section{\label{Mass splittings of the hybrid baryons}
     Masses and mass splittings of the hybrid baryons}

In this section we determine the masses and mass splittings of the flavor
octet and decuplet hybrid baryons which contain two light quarks and
one heavy quark $i$. If the hyperfine interaction $V_{hyp}$ 
(\ref{the interaction Hamiltonian}) would be absent, the $^{spin} flavor$ 
states $\bf^21$, $\bf^28$, $\bf^48$ and $\bf^210$ which form the {\bf 816} 
in eq. (\ref{SU(18) decomposition}), would be degenerate and would have the 
common mass $E_{0i}$. If $V_{hyp}$ is present, the mass of each 
hybrid is given by
\begin{eqnarray}   \label{asolute hybrid mass}
   E_i = E_{0i} + E_{hyp}. 
\end{eqnarray}
We neglect a possible mixing of the hybrids with other states which 
carry the same quantum numbers.

In section \ref{Mass splittings of the flavor singlet hybrid baryons} 
we determine $\kappa_g$ and $\kappa_{ig}$ from the splittings
of the flavor singlet hybrids and calculate $E_{0i}$. In sections 
\ref{Mass splittings of the flavor octet hybrid baryons} and 
\ref{Mass splittings of the flavor decuplet hybrid baryons}
we give the  absolute masses $E_i$ (\ref{asolute hybrid mass}) for the 
flavor octet and decuplet hybrid baryons.

\subsection{\label{Mass splittings of the flavor singlet hybrid baryons}
                   Mass splittings of the flavor singlet hybrid baryons}
                
The effective couplings $\kappa_{ig}$ can be determined from the 
mass splittings of the flavor singlet hybrid baryons.                
We briefly discuss the expected spectrum of the hybrids (compare section
\ref{Wave functions of Hybrids}). The three quarks have to form a flavor 
singlet and a color octet. So their spin state has to be a doublet.  When 
the $\bf^21$ three quark state couples to the constituent 
gluon (spin triplet), a $J=\frac{1}{2}$ and a $J=\frac{3}{2}$ hybrid state
are formed. From eqn. \ref{qq interaction},
eqn. \ref{quark-gluon decomposition} and tables 
\ref{The quark-quark interactions for color octets} 
and \ref{Values for sumiSiSgFiFg} we find: 
\begin{eqnarray} \label{singlet Ehyp}
      E_{hyp} &=& -\frac{7}{8} \kappa_3 -\frac{1}{2}(\kappa-\kappa_3)
                  -\kappa_{3g} \left\{ { -1 \atop 2} \right\}
                     { J=\frac{3}{2} \atop J=\frac{1}{2}} 
\end{eqnarray}
The mass separation is
\begin{eqnarray}
 E_{hyp}(J=3/2)- E_{hyp}(J=1/2)&=& 3  \,\kappa_{3g}.
\end{eqnarray}
For the strange and the charm system we have
\begin{eqnarray}
  \Lambda_s(1520)-\Lambda_s(1405) &=& 115 \,MeV = 3 \, \kappa_{sg}\\
  \Lambda_c(2625)-\Lambda_c(2593) &=&  32 \,MeV = 3 \, \kappa_{cg}.
\end{eqnarray}
So the values for $\kappa_{sg}$ and $\kappa_{cg}$ are determined:
\begin{eqnarray}
 \kappa_{sg} &=& 38 \,MeV \\
 \kappa_{cg} &=& 11 \,MeV.
\end{eqnarray}

We can estimate the mass separation between the
$\Lambda_b(J=3/2)$ and the $\Lambda_b(J=1/2)$. We use 
(\ref{kappa mass relation}) and (\ref{kappa relation})
\begin{eqnarray}
   \kappa_{bg}= \frac{m_c}{m_b} \kappa_{cg}
\end{eqnarray}
to find an estimate value for $\kappa_{bg}$. We use the masses
$m_c=(1.25 \pm 0.15)$  GeV and $m_b=(4.25 \pm 0.15)$ 
GeV \cite{1.0} and find $\kappa_{bg}=(3 \pm 0.5)$ MeV.
It follows
\begin{eqnarray}
   \Lambda_b(J=3/2)-\Lambda_b(J=1/2) &=&  (9 \pm 2) \,MeV.
\end{eqnarray}
Using these values for $\kappa_{sg}$, $\kappa_{cg}$ and  
the relation from eqn. \ref{kappa relation}
we find for $i=s$ and $i=c$ respectively 
\begin{eqnarray}
   \begin{array}{lcccrcr}
  \kappa_g &=& \kappa_{sg}  \frac{\kappa}{\kappa_s}
           &=& 38 \frac{293}{182} \, MeV
           &=& 61 \, MeV, \\
  \kappa_g &=& \kappa_{cg}  \frac{\kappa}{\kappa_c}
           &=& 11 \frac{293}{60} \, MeV
           &=& 54 \, MeV. \\
    \end{array}
\end{eqnarray}
This corresponds to a value of $ \kappa_g$ 
(with a relative error less than 10 $\%$) of
\begin{eqnarray} \label{value of kappag}
 \kappa_g &=&  (58 \pm 4) \, MeV.
\end{eqnarray}
The error of $\kappa_{g}$ is sufficient for the predictive power 
of our effective model. Having gained all values for the various 
$\kappa_{i}$, we can determine $E_{hyp}$ for 
$\Lambda_s(E_s=1405)$ and $\Lambda_c(E_c=2593)$ given in 
(\ref{singlet Ehyp}):
\begin{eqnarray}
E_{hyp}(\Lambda_s)=-291\;MeV; \;\;\;\;\;\;E_{hyp}(\Lambda_c)=-191\;MeV.
\end{eqnarray}
From (\ref{asolute hybrid mass}) we find 
\begin{eqnarray}
   E_{0s}= 1696\; MeV; \;\;\;\;\;\; E_{0c}=2784 \; MeV.
\end{eqnarray}

\subsection{\label{Mass splittings of the flavor octet hybrid baryons}
                   Mass splittings of the flavor octet hybrid baryons}

The three quarks have to form a flavor and a color octet. 
So their spin state may be a doublet or a quartet. There are the 
 $^{spin}flavor$ states $\bf^28$ and $\bf^48$, respectively. When the 
$\bf^28$ three quark state couples to the constituent 
gluon (spin triplet), a $J=\frac{1}{2}$ and a $J=\frac{3}{2}$ hybrid state
is formed. When the $\bf^48$ three quark state couples to the constituent 
gluon, a $J=\frac{1}{2}$,  $J=\frac{3}{2}$ and a $J=\frac{5}{2}$ hybrid state
is formed.

\subsubsection{\label{Splittings in the 
                                {spin}flavor = 28 sector}
                  Splittings in the  
                          $\bf ^{spin}flavor = ^28$ sector}
                          
 From eqn. \ref{qq interaction},
eqn. \ref{quark-gluon decomposition} and tables 
\ref{The quark-quark interactions for color octets} 
and \ref{Values for sumiSiSgFiFg}
we find: 
\begin{eqnarray}
 E_{hyp}(I=1) &=& -\frac{1}{8} \kappa_3
                   +\frac{1}{6} (\kappa -\kappa_3) 
                  -\kappa_{3g} \left\{ { -1/2 \atop 1} \right\}
       -2(\kappa_g-\kappa_{3g})\left\{ { -1/4 \atop 1/2} \right\}
                     { J=\frac{3}{2} \atop J=\frac{1}{2}} \\
 E_{hyp}(I=0) &=& -\frac{1}{8} \kappa_3
                  -\frac{1}{2} (\kappa -\kappa_3) 
                  -\kappa_{3g} \left\{ { -1/2 \atop 1} \right\}
       -2(\kappa_g-\kappa_{3g})\left\{ { 0 \atop 0} \right\}
                     { J=\frac{3}{2} \atop J=\frac{1}{2}}                    
\end{eqnarray}                       
We list the resulting absolute masses $E_i$ 
(\ref{asolute hybrid mass}) in 
table \ref{Masses of the flavor decuplet and flavor octet hybrids}.

\subsubsection{\label{Splittings in the 
                                 {spin}flavor = 48 sector}
                   Splittings in the 
                           $\bf ^{spin}flavor = ^48$ sector} 
                
 From eqn. \ref{qq interaction},
eqn. \ref{quark-gluon decomposition} and tables 
\ref{The quark-quark interactions for color octets} 
and \ref{Values for sumiSiSgFiFg}
we find: 
\begin{eqnarray}
  E_{hyp}(I=1) = +\frac{1}{8} \kappa_3
                   +\frac{1}{6} (\kappa -\kappa_3) 
                  -\kappa_{3g} \left\{ 
                      \begin{array}{c} -3/2 \\ 1 \\ 5/2   
                      \end{array}    
                  \right\} 
       -2(\kappa_g-\kappa_{3g})\left\{ 
                      \begin{array}{c} -3/8 \\ 1/4 \\ 5/8   
                      \end{array}    
                  \right\}
                    \begin{array}{c} J=5/2 \\ J=3/2 \\ J=1/2  
                    \end{array}\\
 E_{hyp}(I=0) = +\frac{1}{8} \kappa_3
                  -\frac{1}{12} (\kappa -\kappa_3) 
                  -\kappa_{3g} \left\{ 
                      \begin{array}{c} -3/2 \\ 1 \\ 5/2   
                      \end{array}    
                  \right\} 
       -2(\kappa_g-\kappa_{3g})\left\{ 
                      \begin{array}{c} -5/8 \\ 5/12 \\ 25/24   
                      \end{array}    
                  \right\}
                    \begin{array}{c} J=5/2 \\ J=3/2 \\ J=1/2  
                    \end{array}\\
\end{eqnarray}                          
We list the resulting absolute masses $E_i$ 
(\ref{asolute hybrid mass}) in 
table \ref{Masses of the flavor decuplet and flavor octet hybrids}.

\subsection{\label{Mass splittings of the flavor decuplet hybrid baryons}
                   Mass splittings of the flavor decuplet hybrid baryons}

The three quarks form a flavor decuplet and a color octet. 
So their spin state must be a doublet to ensure antisymmetry 
under permutations. When the $\bf^2 10$ three quark state couples to the 
constituent gluon (spin triplet), a $J=\frac{1}{2}$ and a $J=\frac{3}{2}$ 
state are formed. From eqn. \ref{qq interaction},
eqn. \ref{quark-gluon decomposition} and tables 
\ref{The quark-quark interactions for color octets} 
and \ref{Values for sumiSiSgFiFg}
we find: 
\begin{eqnarray}
 E_{hyp}(I=1) &=& +\frac{5}{8} \kappa_3
                  +\frac{1}{6} (\kappa -\kappa_3) 
                  -\kappa_{3g} \left\{ { 0 \atop 0} \right\}
       -2(\kappa_g-\kappa_{3g})\left\{ { -1/4 \atop 1/2} \right\}
                     { J=\frac{3}{2} \atop J=\frac{1}{2}}                  
\end{eqnarray}                   
We list the resulting absolute masses $E_i$ 
(\ref{asolute hybrid mass}) in 
table \ref{Masses of the flavor decuplet and flavor octet hybrids}.                          
(We also give the mass splittings $\Delta E_b$ for all the beauty 
hybrids).                          
                                                    
\begin{table}[h]
 \[
   \begin{tabular}{|c|c|c|c|c|c|} \hline
   $^{spin}flavor$& $J^P$ & $E_s/MeV$ & $E_c/MeV$& $\Delta E_b/MeV$ \\ \hline   
$\bf^48$& $5/2^-$ & 1809 & 2882 & $E(5/2)-E(3/2) = 76$ \\ 
 (I=1)  & $3/2^-$ & 1689 & 2796 & $E(3/2)-E(1/2) = 46$  \\ 
        & $1/2^-$ & 1617 & 2744 &                       \\ \hline   
$\bf^48$& $5/2^-$ & 1792 & 2847 & $E(5/2)-E(3/2) = 123$ \\ 
 (I=0)  & $3/2^-$ & 1655 & 2722 & $E(3/2)-E(1/2) = 73$  \\ 
        & $1/2^-$ & 1573 & 2647 &                       \\ \hline
$\bf^28$& $3/2^-$ & 1721 & 2844 & $E(3/2)-E(1/2) = 87$ \\ 
 (I=1)  & $1/2^-$ & 1634 & 2757 &                       \\ \hline               
$\bf^28$& $3/2^-$ & 1637 & 2666 & $E(3/2)-E(1/2) = 5$ \\ 
 (I=0)  & $1/2^-$ & 1580 & 2649 &                       \\ \hline
$\bf^210$&$3/2^-$ & 1838 & 2884 & $E(3/2)-E(1/2) = 83$  \\ 
 (I=1)  & $1/2^-$ & 1808 & 2813 &                       \\ \hline
   \end{tabular}
  \]
 \caption{\label{Masses of the flavor decuplet and flavor octet hybrids}
                 Masses of the flavor decuplet and flavor octet hybrids}
\end{table}

\section{\label{The parity of Lambda(1405)}
                The parity of $\Lambda(1405)$}
                
We assumed so far, that the parity of $\Lambda(1405)$ is negative as
listed in PDG \cite{1.0}. This choice is made because in the quark
model, the $\Lambda(1405)$ necessarily needs to be orbitally excited
and thus must have negative parity.  However, there is no direct
evidence from experiment that $\Lambda(1405)$ actually has negative
parity. Hemingway \cite{Hem} writes, that a ``Byers-Fenster spin-parity
analysis gives no parity discrimination". Thomas \cite{Thom} writes,
that ``the experimental facts are that the parity of $\Lambda(1405)$
has not yet been determined in a production experiment" and that they
were ``unable to make a parity determination".  We therefore regard
the parity of the $\Lambda(1405)   $ as experimentally undetermined.
If the $\Lambda(1405)$ is the lightest hybrid baryon, the hybrid model
strongly suggests it actually has even parity because the bag model
predicts \cite{Closebook} the lightest
$J^P=1^{(-)} (TM)$ gluon mode is about 300 MeV heavier than the lightest
$J^P=1^{(+)} (TE)$ gluon mode.  The negative parity partner of the
$\Lambda(1405)$ hybrid would thus be about 300 MeV heavier with mass
of about 1.7 GeV. It's experimental detection would be more difficult
due to mixing with other states.  A second possibility is that the
$\Lambda(1405)$ has odd parity and there is
a lower mass pair of even parity states. If the shift is
about 300 MeV as predicted from bag models, there would be $J^P=\frac{1}{2}^{(+)},
\frac{3}{2}^{(+)}$ states with strangeness -1 and mass about 1.1 and 1.2
GeV. It is doubtful to us that such low energy states could be discovered 
by partial wave analysis. They would be far below threshold in the
$\Sigma$ $\pi$ channel.  A positive parity $\Lambda(1100)$ would
only impact the L=1 state and furthermore would have little impact
compared to the $\Lambda(1405)$ which is much closer to the physical
region. 

\section{\label{A low lying dihyperon ?}
                A low lying dihyperon ?}

We are exploring the ansatz that a $uds$ in a color octet, flavor
singlet state binds with a constituent gluon to produce the
$\Lambda(1405)$.  We analyzed mass splittings between members of the
various multiplets, but we made no absolute mass predictions.  (Those
are model dependent and would require the use of a model such as the
Skyrme, MIT bag or potential models). However, if we make the hybrid
baryon ansatz for these states, we know the masses experimentally.
Lattice calculations \cite{glue} give the lightest glueball in the
range 1.4 - 1.7 GeV, where there are good glueball
candidates.  Thus if the hybrid baryon ansatz is correct, the
approximate coincidence of the $udsg$ and $gg$ masses suggests that a
$uds$ in a flavor singlet color octet state is approximately
equivalent to a gluon from the dynamical point of view.  The dynamics
of a hadronic bound state depends primarily on the color, mass and
spin of the constituents.  Thus a spatially-compact $uds$ system in a
flavor singlet state would behave like a gluon with spin 1/2 rather
than spin 1. Thus the dynamical similarity of the $uds$ and gluon can
only be approximate and mass estimates must have about a 100 MeV
uncertainty at least.  Making this ansatz, a combination of two $uds$
should be a glueball-like state with mass also of about 1.5 GeV. This
would be the $H$ dihyperon, which is a six quark state with total spin
and isospin zero, baryon number 2 and strangeness -2. The dihyperon
was predicted in 1977 by Jaffe \cite{Jaffe} in a MIT bag model
calculation with mass of 2150 MeV.  Since that, many other dihyperon
mass calculations have been performed, also using Skyrme
\cite{skyrme}, \cite{skyrme2}
and quark cluster models \cite{cluster}.  The mass estimates for the
lowest lying dibaryon H range from 1.5 to 2.2 GeV. The differences in
the mass predictions are attributed to the difference between the
models which are characterized by model parameters and the model
dependent assumptions which are made in order to describe hadronically
bound states.  If the $\Lambda(1405)$ is a hybrid baryon, we therefore
suspect that the mass of the $H$ dibaryon is in the 1.5 GeV mass
region, as references \cite{skyrme} suggest.  We leave
to another work a discussion of the phenomenological issues and
detectability of such a light $H$.

\section{\label{Summary and Conclusion}
                Summary and conclusion}

We have explored the hypothesis that four particles $\Lambda_s(1405)$,
$\Lambda_s(1520)$, $\Lambda_c(2593)$ and $\Lambda_c(2676)$, are
hybrids.  The observed mass splittings are consistent with the hybrid
baryon hypothesis, resolving a severe problem of the conventional
identification as an orbital excitation of a 3 quark state.  It is
non-trivial that the ordering of states is $m_{J=3/2}>m_{J=1/2}$, as
observed experimentally; in the conventional $L=1$ picture the
spin-3/2 state is necessarily the lightest.  Assuming these states are
flavor singlet hybrids fixes the parameters of the quark-gluon
hyperfine interaction.  This allows the mass splittings
of the flavor octet and decuplet hybrid baryons to be predicted, but
without developing a theory of mixing with nearby ordinary octets and
decuplets these predictions cannot be tested.

The best test of the ansatz that the $\Lambda_s(1405)$,
$\Lambda_s(1520)$, $\Lambda_c(2593)$ and $\Lambda_c(2676)$, are
hybrids is that they will be parity doubled, with the odd parity
partner about 300 MeV heavier than the even parity state.  Thus we
predict either that the $\Lambda_s(1405)$, $\Lambda_s(1520)$,
$\Lambda_c(2593)$ and $\Lambda_c(2676)$ are even parity, or that there
are as-yet-undiscovered even parity flavor singlet, strangeness -1
states at about 1.1 and 1.2 GeV.  The hybrid ansatz suggests, but does
not predict, that the $H$ dibaryon mass is around 1.5 GeV.

\section{Acknowledgement}

O. K. was supported by the \emph{Rutgers University} and  
\emph{DAAD ("Deutscher Akademischer Austauschdienst")}. O. K. is grateful to
G. T. Gabadadze, C. Roemelsberger, F. v. d. Pahlen and I. Paul for useful 
discussions.

\appendix

\section{\label{The Pauli and the Gell-Mann matrices}
                The Pauli and the Gell-Mann matrices}

The Pauli matrices \cite{Closebook} p. 23 are chosen to be
\[
  \sigma_1=
  \left(
        \begin{array}{rr}
         0&1\\1&0
        \end{array}
  \right),\;\;\;\;
  \sigma_2=
  \left(
        \begin{array}{rr}
         0&-i\\i&0
        \end{array}
  \right),\;\;\;\;
  \sigma_3=
  \left(
        \begin{array}{rr}
         1&0\\0&-1
        \end{array}
  \right)
\]
The Gell-Mann matrices \cite{Closebook} p. 30 are given by
\begin{eqnarray*}
  \lambda_1=
       \left(
             \begin{array}{rrr}
              \;\;\; 0 & 1 &0 \\ 1 &\;\;\; 0 & 0 \\0 & 0 &\;\;\; 0
             \end{array}
       \right)
&
\lambda_2=
       \left(
             \begin{array}{rrr}
               0 &-i &0 \\ \;\;\; i & 0 & 0 \\0 & 0 &\;\;\; 0
             \end{array}
       \right)
&
\lambda_3=
       \left(
             \begin{array}{rrr}
              \;\;\; 1 & 0 & 0 \\ 0 &-1 & 0 \\0 & 0 &\;\;\; 0
             \end{array}
       \right)
\\
 \lambda_4=
       \left(
             \begin{array}{rrr}
              \;\;\; 0 & 0 & 1 \\0 &\;\;\; 0 & 0 \\1 & 0 &\;\;\; 0
             \end{array}
       \right)
&
    \lambda_5=
       \left(
             \begin{array}{rrr}
               0 & 0 &-i \\ \;\;\; 0 & 0 & 0 \\i & \;\;\;0 & 0
             \end{array}
       \right)
&
    \lambda_6=
       \left(
             \begin{array}{rrr}
               0 & \;\;\;0 &0 \\\;\;\; 0 & 0 & 1 \\0 & 1 &\;\;\; 0
             \end{array}
       \right)
\\
   \lambda_7=
       \left(
             \begin{array}{rrr}
             \;\;\;  0 & 0 & 0 \\ 0 &\;\;\; 0 & -i \\ 0 & i & 0
             \end{array}
       \right)
&
    \lambda_8= \displaystyle
       \frac{1}{\sqrt{3}}
       \left(
             \begin{array}{rrr}
              \;\;\; 1 & 0 & 0 \\ 0 &\;\;\; 1 & 0 \\ 0 & 0 & -2
             \end{array}
       \right)&
\end{eqnarray*}

\section{\label{Casimir operators}
                Casimir operators}

The Casimir operators $C_{2,3}$ \cite{Li} pp. 88, 89 are defined:
\begin{eqnarray}
   SU(2): &C_2& = S\cdot S =\sum_{m=1}^3 S_mS_m =
                  \frac{1}{4}\sum_{m=1}^3 \sigma_m\sigma_m \\
          &S_m& =   \frac{1}{2}\sigma_m \\
   SU(3): &C_3& = F\cdot F =\sum_{a=1}^8 F_aF_a =
                   \frac{1}{4}\sum_{a=1}^8 \lambda_a\lambda_a \\ 
          &F_a& =  \frac{1}{2}\lambda_a 
\end{eqnarray}
The values of these operators in various representations are given in
table \ref{Values for $SU(2)$ and $SU(3)$ Casimir operators in
various representations}. We use $C_{2}=S\cdot (S+1)$ . 
The values for $C_{3}$ can be found in \cite{Ca} p. 74. 
\begin{table}[h]
 \[
  SU(2):  
           \begin{array}{|c||c|c|c|c|c|} \hline
            dim &{\bf1}&{\bf2}&{\bf3}&{\bf4 }&{\bf5} \\ \hline 
            C_2 &  0   &  3/4 &  2   &  15/4 &     6 \\ \hline     
           \end{array}
          \;\;\;\;\;
   SU(3):     
           \begin{array}{|c||c|c|c|c|c|} \hline
             dim &{\bf1}&{\bf3}&{\bf6}&{\bf8}& {\bf10} \\ \hline   
             C_3 &  0   &  4/3 & 10/3 &   3  &  6      \\ \hline
           \end{array}
   \]   
  \caption{\label{Values for $SU(2)$ and $SU(3)$ Casimir operators in
                     various representations}
                  Values for $SU(2)$ and $SU(3)$ Casimir operators in
                     various representations}
\end{table}

\section{\label{Definitions of the mixed symmetry functions}
                   Definitions of the mixed symmetry functions}

The $\varphi_{MS}$ and  $\varphi_{MA}$ are defined in the table
\ref{The functions} .
This table is taken from \cite{Closebook} p. 46. 
These states are mixed symmetric or  mixed antisymmetric. 
They are symmetric or antisymmetric under interchange of the first 
two quarks. The various wave functions for $SU(3)$ are named by the 
name of the particle they belong to.
\begin{table}[h]
 \begin{tabular}{|c|c|c|} \hline
  $label$  &  $\varphi_{MS}   $   &  $\varphi_{MA}  $ \\ \hline \hline
  $P$  &  $   \frac{1}{\sqrt{6}}
             \left[ (ud+du)u -2uud
             \right]$   
        &  $\frac{1}{\sqrt{2}}(ud-du)u  $ \\ \hline
 $N$  &  $  - \frac{1}{\sqrt{6}}
             \left[ (ud+du)d -2ddu
             \right]$   
 &  $\frac{1}{\sqrt{2}}(ud-du)d  $ \\ \hline
 $\Sigma^+$ &  $   \frac{1}{\sqrt{6}}
             \left[ (us+su)u -2uus
             \right]$   
&  $\frac{1}{\sqrt{2}}(us-su)u  $ \\ \hline
 $\Sigma^0$ &$  \frac{1}{\sqrt{6}}
           \left[ s \left( \frac{du+ud}{\sqrt{2}} \right)
                  + \left( \frac{dsu+usd}{\sqrt{2}} \right)
          \right. $
 &  $\frac{1}{\sqrt{2}}
          \left[   \left( \frac{dsu+usd}{\sqrt{2}} \right)
                -s \left( \frac{ud+du}{\sqrt{2}}  \right)
            \right]$ \\
 & $ \left.  
          -2 \left( \frac{du+ud}{\sqrt{2}} \right)s  
  \right]$ &  \\ \hline
$\Sigma^-$ &  $  \frac{1}{\sqrt{6}}
             \left[ (ds+sd)d -2dds
             \right]$   
&  $\frac{1}{\sqrt{2}}(ds-sd)d  $ \\ \hline
$\Lambda^0$ &  $\frac{1}{\sqrt{2}}
          \left[   \left( \frac{dsu-usd}{\sqrt{2}} \right)
                +         \frac{s(du-ud)}{\sqrt{2}}  
         \right]$
 &$   \frac{1}{\sqrt{6}} 
         \left[     \frac{s(du-ud)}{\sqrt{2}} 
                +   \frac{usd-dsu}{\sqrt{2}}  
         \right.   $ \\ 
&&$ \left. -\frac{2(du-ud)s}{\sqrt{2}} \right]$ \\ \hline
$\Xi^-$ &  $ - \frac{1}{\sqrt{6}}
             \left[ (ds+sd)s -2ssd
             \right]$   
 &  $\frac{1}{\sqrt{2}}(ds-sd)s  $ \\ \hline
$\Xi^0$ &   $ - \frac{1}{\sqrt{6}}
             \left[ (us+su)s -2ssu
             \right]$   
 &  $\frac{1}{\sqrt{2}}(us-su)s  $ \\ \hline
 \end{tabular}
\caption{\label{The functions}The functions of mixed symmetry}
\end{table}

\section{\label{Sample calculation for the flavor singlet}
                Sample calculation for the flavor singlet}

In this appendix we give an example of calculating the 
first term of $V_{qg}$ in eqn. \ref{quark-gluon decomposition}.
To evaluate the first term, we need to look at the structure of the 
completely antisymmetric three quark wave function,
discussed in section \ref{Wave functions of Hybrids}.
It follows for the  flavor singlet state $\bf^21$:
 \begin{eqnarray}
    \lefteqn{ <\sum_i S^i\cdot S^gF^i\cdot F^g >= 
               3  <S^3\cdot S^gF^3\cdot F^g>=
             } &&\nonumber \\  
   &=&  \frac{3}{2}
                  \Big\{  
                     <S^3\cdot S^g>_{MA}<F^3\cdot F^g>_{MA} +
                     <S^3\cdot S^g>_{MS}<F^3\cdot F^g>_{MS}
                  \Big\}.
  \end{eqnarray}
(The first line follows from the antisymmetry.)
The subscripts $MA$ and $MS$, introduced in section 
\ref{Construction of the quark wave functions}, indicate that 
the pair of the first and second quark is 
an antisymmetric or a symmetric state, respectively. 
In the color case this means for $<F^3\cdot F^g>_{MA}$ that the first and
second quark are in a color antitriplet, $\bf\bar{3}$. So the
gluon-third quark system has to be in a color triplet to form an overall
singlet with the spectator diquark. Similarly 
for $<F^3\cdot F^g>_{MS}$, the
first and second quark form a symmetric state, the color sextet. 
Thus the gluon and the third quark are an antisextet.
For the flavor singlet with total $J =1/2$ we have for the color operator:
 \begin{eqnarray}  
     \label{color gluon third quark}
     < F^{3}\cdot F^{g}>_{MA}^{MS} &=&  \frac{1}{2} 
                       \Big[        
                            (F^{3}+F^{g})^2
                           -(F^{3})^2-(F^{g})^2 
                        \Big]   \\     
                   &=& \frac{1}{2} 
                   \Big[ 
                         \left\{ 
                                {
                                  C^{3g}_{F} \{ {\bf \bar{6}} \}  \atop
                                  C^{3g}_{F} \{{\bf  3 } \}
                                }
                        \right\}
                              - C^{q}_{F} \{ {\bf3} \} 
                              - C^{g}_{F} \{ {\bf8} \}
                   \Big] \\
                   &=& \frac{1}{2} 
                   \left[ 
                         \left\{ 
                                {
                                 10/3  \atop 4/3 
                                } 
                        \right\}
                             -  \frac{4}{3}  
                             -  3
                  \right]  
                               = 
                    \left\{ 
                             {-1/2  \atop -3/2} 
                    \right\}.
  \end{eqnarray}
In the spin case, we have multiple possibilities for the constituent 
gluon (spin 1) to couple to the three quark system (spin 1/2) to 
give an overall spin $1/2$ $ qqqg$-state. $<S^3\cdot S^g>_{MA}$ is easy
to evaluate because the diquark is in an antisymmetric state, i.e., it 
has spin 0. Thus the gluon and the third quark carry the total spin 
of the hadron, 1/2.
  \begin{eqnarray}  
     \label{ms gluon third quark}
     < S^{3}\cdot S^{g}>_{MA} &=&  \frac{1}{2} 
                       \Big[        
                            (S^{3}+S^{g})^2
                           -(S^{3})^2-(S^{g})^2 
                        \Big]   \\     
                   &=& \frac{1}{2} 
                   \left[ 
                         \left\{ 
                                {
                                  C^{3g}_{S} \{ {\bf4} \}  \atop
                                  C^{3g}_{S} \{ {\bf2} \}
                                }
                        \right\}
                              - C^{q}_{S} \{ {\bf2} \} 
                              - C^{g}_{S} \{ {\bf3} \}
                   \right] \\
                   &=& \frac{1}{2} 
                   \left[ 
                         \left\{ 
                                {
                                 15/4  \atop 3/4 
                                } 
                        \right\}
                             -  \frac{3}{4}  
                             -  \frac{8}{4}
                  \right]  
                               = 
                    \left\{ 
                             {1/2  \atop -1} 
                    \right\}.
  \end{eqnarray}
The determination of  the $<S^3\cdot S^g>_{MS}$ value is more delicate. 
The diquark is in a spin 1 state, so the gluon and the
third quark can be either in an $S=1/2$ or $S=3/2$ state to couple
with the diquark to an overall $qqqg$ having spin 1/2. 
We have
to rewrite the $qqqg$-state in terms of gluon-third quark states in
order to evaluate the expectation value in eqn. \ref{ms gluon third
quark}, namely we have to
see how often $|gq>_{S=1/2}$ and $|gq>_{S=3/2}$ are involved. This is
done with the help of Clebsch-Gordan coefficients.
We use the notation $|l,m>_{X}$, where $l$ labels the spin
representation, $m$ is the projection on the $z$-axis and $X$
indicates which particles form that state. All Clebsch-Gordan coefficients
are taken from the Particle Data Group book \cite{1.0}.
First we couple the gluon state $|1>_g$ to the mixed symmetric part
of the three quark state $|\frac{1}{2}>_{qqq}$, to get a doublet.
  \begin{eqnarray}
    \begin{array}{rcl}
   | \frac{1}{2}, \frac{1}{2} >_{qqqg} &=& 
          \sqrt{ \frac{2}{3}} |1,1>_g  |\frac{1}{2},-\frac{1}{2}>_{qqq}
         -\sqrt{ \frac{1}{3}} |1,0>_g  |\frac{1}{2}, \frac{1}{2}>_{qqq}
   \end{array}  
 \end{eqnarray}
In this formula we substitute the mixed symmetric
$|\frac{1}{2}>_{qqq}$ states,
   \begin{eqnarray}
        \begin{array}{rcl}
    |\frac{1}{2}, \frac{1}{2}>_{qqq}&=&
     \sqrt{ \frac{2}{3}} |1,1>_{12} |\frac{1}{2},-\frac{1}{2}>_{3}
    -\sqrt{ \frac{1}{3}} |1,0>_{12} |\frac{1}{2}, \frac{1}{2}>_{3}\\
   |\frac{1}{2},-\frac{1}{2}>_{qqq}&=&
     \sqrt{ \frac{1}{3}} |1,0>_{12} |\frac{1}{2},-\frac{1}{2}>_{3}
    -\sqrt{ \frac{2}{3}} |1,-1>_{12}|\frac{1}{2}, \frac{1}{2}>_{3}, 
   \end{array}  
 \end{eqnarray}
and get
  \begin{eqnarray}
    \begin{array}{rcl} \label{The qqqg 1/2 state}
     \lefteqn{ \textstyle | \frac{1}{2}, \frac{1}{2} >_{qqqg}=}\\ 
        &&\sqrt{ \frac{2}{3}} \sqrt{ \frac{1}{3}}
        |1,1>_g  |1,0>_{12}  |\frac{1}{2},-\frac{1}{2}>_{3}
        -\sqrt{ \frac{2}{3}} \sqrt{ \frac{2}{3}}
        |1,1>_g  |1,-1>_{12} |\frac{1}{2}, \frac{1}{2}>_{3} \\
        &&- \sqrt{ \frac{1}{3}} \sqrt{ \frac{2}{3}}
        |1,0>_g  |1,1>_{12} |\frac{1}{2},-\frac{1}{2}>_{3}
       +\sqrt{ \frac{1}{3}}\sqrt{ \frac{1}{3}}
       |1,0>_g |1,0>_{12}|\frac{1}{2}, \frac{1}{2}>_{3}.
    \end{array}  
  \end{eqnarray}
We have to rewrite this in terms of  $|l,m>_{3g}$-states 
with definite values.
 \begin{eqnarray}
  \begin{array}{lcl}
    | \frac{3}{2}, \frac{3}{2} >_{3g} &=&
         |1,1>_g |\frac{1}{2}, \frac{1}{2}>_{3}\\
    | \frac{3}{2}, \frac{1}{2} >_{3g} &=&
     \sqrt{ \frac{1}{3}} |1,1>_g |\frac{1}{2},-\frac{1}{2}>_{3}
    +\sqrt{ \frac{2}{3}} |1,0>_g |\frac{1}{2}, \frac{1}{2}>_{3}\\
    | \frac{3}{2},-\frac{1}{2} >_{3g} &=&
     \sqrt{ \frac{2}{3}} |1,0>_g |\frac{1}{2},-\frac{1}{2}>_{3}
    +\sqrt{ \frac{1}{3}} |1,-1>_g|\frac{1}{2}, \frac{1}{2}>_{3}\\
    | \frac{3}{2},-\frac{3}{2} >_{3g} &=&
         |1,-1>_g |\frac{1}{2},-\frac{1}{2}>_{3}\\
    | \frac{1}{2}, \frac{1}{2} >_{3g} &=&
     \sqrt{ \frac{2}{3}} |1,1>_g |\frac{1}{2},-\frac{1}{2}>_{3}
    -\sqrt{ \frac{1}{3}} |1,0>_g |\frac{1}{2}, \frac{1}{2}>_{3}\\
    | \frac{1}{2},-\frac{1}{2} >_{3g} &=&
     \sqrt{ \frac{1}{3}} |1,0>_g |\frac{1}{2},-\frac{1}{2}>_{3}
    -\sqrt{ \frac{2}{3}} |1,-1>_g|\frac{1}{2}, \frac{1}{2}>_{3}
  \end{array}
\end{eqnarray}
Inverting these equations we find:
\begin{eqnarray}
   \begin{array}{rcl} \label{Gluon third quark states}
        |1,1>_g |\frac{1}{2},-\frac{1}{2}>_{3} &=&
        \sqrt{ \frac{1}{3}} | \frac{3}{2}, \frac{1}{2} >_{3g} 
       +\sqrt{ \frac{2}{3}} | \frac{1}{2}, \frac{1}{2} >_{3g}.\\
        |1,0>_g |\frac{1}{2},-\frac{1}{2}>_{3} &=&
        \sqrt{ \frac{2}{3}} | \frac{3}{2},-\frac{1}{2} >_{3g} 
      + \sqrt{ \frac{1}{3}} | \frac{1}{2},-\frac{1}{2} >_{3g}.\\
        |1,0>_g |\frac{1}{2}, \frac{1}{2}>_{3} &=&
        \sqrt{ \frac{2}{3}}| \frac{3}{2},\frac{1}{2} >_{3g} 
        -\frac{1}{\sqrt{3}}| \frac{1}{2}, \frac{1}{2} >_{3g}.
    \end{array}
 \end{eqnarray}
We substitute these results in  eqn. \ref{The qqqg 1/2 state}
and find:
\begin{eqnarray}
  \begin{array}{rcl}
    | \frac{1}{2}, \frac{1}{2} >_{qqqg} &=& 
       - \frac{2}{3} 
       |1,-1>_{12} | \frac{3}{2}, \frac{3}{2} >_{3g}\\
 &&    + \frac{2}{3} \sqrt{\frac{2}{3}}
       |1,0>_{12}| \frac{3}{2}, \frac{1}{2} >_{3g} \\
 &&    -\frac{2}{3\sqrt{3}}
       |1,1>_{12}| \frac{3}{2},-\frac{1}{2} >_{3g} \\
 &&    + \frac{1}{3\sqrt{3}}
       |1,0>_{12}| \frac{1}{2}, \frac{1}{2} >_{3g} \\
 &&    -\frac{\sqrt{2}}{3\sqrt{3}}
       |1,1>_{12}| \frac{1}{2},-\frac{1}{2} >_{3g}
   \end{array}
 \end{eqnarray}
From this follows
  \begin{eqnarray}
  <S^3\cdot S^g>_{MS}&=&
   \frac{12+8+4}{27}<S^3\cdot S^g>_{MS}^{S=3/2}
   + \frac{1+2}{27}<S^3\cdot S^g>_{MS}^{S=1/2} \\
   &=&
   \frac{24}{27}(\frac{1}{2})+
   \frac{3}{27}(-1) = \frac{9}{27}= \frac{1}{3}.
  \end{eqnarray}
With these values, $<\sum_i S^i\cdot S^gF^i\cdot F^g>$in the total
spin 1/2 state is 
   \begin{eqnarray}
      \lefteqn{
                <\sum_i S^i\cdot S^gF^i\cdot F^g> =  
              } \nonumber \\
    &&=\frac{3}{2}\left\{  
                      <S^3\cdot S^g>_{MA}<F^3\cdot F^g>_{MA} +
                      <S^3\cdot S^g>_{MS}<F^3\cdot F^g>_{MS}
                  \right\}\\
    &&= \frac{3}{2}
                \left\{ 
                  (-1 )(-\frac{3}{2} )+
                  (\frac{1}{3} )(-\frac{1}{2} )
                \right\}
      = \frac{3}{2}
                \left\{ 
                  \frac{9}{6}-\frac{1}{6} 
                \right\} \\
       && =2.
    \end{eqnarray}

\end{document}